# Signature of Superconductivity in Pressurized Trilayer-nickelate Pr$_4$Ni$_3$O$_{10-\delta}$

Xing Huang[1#], Hengyuan Zhang[1#], Jingyuan Li[1], Mengwu Huo[1], Junfeng Chen[1], Zhengyang Qiu[1], Peiyue Ma[1], Chaoxin Huang[1], Hualei Sun[2*], Meng Wang[1*]

[1]*Center for Neutron Science and Technology, Guangdong Provincial Key Laboratory of Magnetoelectric Physics and Devices, School of Physics, Sun Yat-Sen University, Guangzhou, Guangdong 510275, China*
[2]*School of Sciences, Sun Yat-Sen University, Shenzhen, Guangdong 518107, China*
# *These authors contribute equally to this work.*
*sunhlei@mail.sysu.edu.cn (H. L. Sun), wangmeng5@mail.sysu.eud.cn (M. Wang)

**Abstract**

The discovery of high-temperature superconductivity in La$_3$Ni$_2$O$_7$ and La$_4$Ni$_3$O$_{10}$ under pressure has drawn extensive attention. Herein, we report systematic investigations on the evolutions of structure, magnetism, and electrical resistance of Pr$_4$Ni$_3$O$_{10-\delta}$ polycrystalline samples under various pressures. Pr$_4$Ni$_3$O$_{10-\delta}$ exhibits density wave transitions on Ni and Pr sublattices at about 158 K and 4.3 K, respectively, and the density wave can be progressively suppressed by pressure. A structural transformation from the monoclinic $P2_1/a$ space group to the tetragonal $I4/mmm$ occurs at around 20 GPa. An apparent drop in resistance with evident magnetic field dependence is observed as pressure above 20 GPa, indicating the emergence of superconductivity in Pr$_4$Ni$_3$O$_{10-\delta}$ polycrystalline samples. The discovery of the signature of superconductivity in Pr$_4$Ni$_3$O$_{10-\delta}$ broadens the family of nickelate superconductors and provides a new platform for investigating the mechanisms of superconductivity in the Ruddlesden–Popper phases of nickelates.

## 1 Introduction

The similarity between the electronic configurations of Ni$^+$ ($3d^9$) and Cu$^{2+}$ ($3d^9$) hints at a great potential to realize high transition temperature (HT$_c$) superconductivity in nickelates[1]. After more than two decades of effort, the superconductivity with $T_c$ around 9–15 K was indeed observed in hole-doped infinite layer Nd$_{1-x}$Sr$_x$NiO$_2$ thin films[2–5]. Very recently, the superconducting transition temperature ($T_c$) has been improved to 32 K in Sm$_{0.79}$Eu$_{0.12}$Ca$_{0.04}$Sr$_{0.05}$NiO$_2$ thin film[6]. The doped infinite-layer (112) nickelates can be obtained by soft-chemistry topotactic reduction from the corresponding Ruddlesden–Popper (RP) phase $R$NiO$_3$[7–10]. The valence state of Ni is +1.2 in superconducting 112 compounds; the number of $3d$ electrons is the same as that of cuprate HT$_c$ superconductors. An exciting breakthrough was the discovery of HT$_c$ superconductivity with $T_c$=80 K in pressurized bulk samples of bilayer RP nickelate La$_3$Ni$_2$O$_7$, where the Ni shows a nominal mixed valence state of Ni$^{2.5+}$ ($3d^{7.5}$)[11–16]. The superconductivity was also observed in Pr doped bilayer RP phase La$_2$PrNi$_2$O$_7$ [17,18]and trilayer PR phase La$_4$Ni$_3$O$_{10}$ under high pressure[19–24]. However, the maximum $T_c$ of the trilayer phase of 30 K is much lower than that of the bilayer phase.

Density wave transitions have been observed in $La_3Ni_2O_7$ and $La_4Ni_3O_{10}$ at ambient pressure. μSR[25,26], NMR[27], and RIXS[28] measurements reveal a spin density wave (SDW) transition in $La_3Ni_2O_7$ with a transition temperature around 150 K[29]. Optical conductivity measurements on $La_3Ni_2O_7$ reveal an energy gap opened below 115 K, suggesting a density wave-like transition related to charge in $La_3Ni_2O_7$[30]. An intertwined SDW and charge density wave (CDW) on the Ni sublattice forms below 135 K in the trilayer nickelate $La_4Ni_3O_{10}$[31,32]. These emergent density waves exhibit a complex relationship with the superconductivity under pressure in nickelates. Searching for new superconducting nickelate compounds is highly desired to elucidate the mechanism of superconductivity in the RP phase of nickelates.

It was suggested theoretically that the $T_c$ of $R_3Ni_2O_7$ ($R$ = rear earth metals) could be dramatically tuned by the $R$ element substitution[33–35]. However, partial substitution of $La^{3+}$ with $Pr^{3+}$ ions does not change the superconductivity, while the superconducting phase purity is improved significantly in $La_2PrNi_2O_7$[17,18]. A higher doping level of the rare earth elements in $La_{3-x}R_xNi_2O_7$ is still challenging. Fortunately, the trilayer nickelates have a large family of compounds, wherein $R_4Ni_3O_{10}$ ($R$ = La, Pr, and Nd) phases are relatively stable[36–38], providing an ideal platform to explore the effect of rare-earth engineering on the superconductivity in the trilayer nickelates.

$Pr_4Ni_3O_{10}$ develops quasi-two-dimensional (2D) intertwined density waves on the Ni sublattice at $T_{DW} \approx 158$ K[39–42]. A distinctly anisotropic behavior can be observed below $T_{DW}$: a metal-to-metal and metal-to-insulator-like transition along the $ab$ plane and $c$ axis, respectively. $Pr^{3+}$ ions build the interlayer exchange pathways to drive a fully 3D-SDW ground state below 5 K and 26 K in single crystal[40] and polycrystalline samples, respectively[42]. The coupling between the local Pr $4f$ sublattice and Ni $3d$ sublattice may result in a distinct scenario from $La_4Ni_3O_{10}$ under pressure. Thus, in this work, we investigate the structure and electronic transport properties of $Pr_4Ni_3O_{10-\delta}$ polycrystalline samples under pressure. We find the monoclinic structure of $Pr_4Ni_3O_{10-\delta}$ is pressurized into tetragonal above ~20 GPa, accompanied by a significant drop in resistance at 30 K. The evolutions of the resistance drop under magnetic field and pressure strongly indicate a superconducting transition as that observed in $La_3Ni_2O_7$.

## 2 Experimental methods

$Pr_4Ni_3O_{10-\delta}$ polycrystalline sample was synthesized using the sol-gel method. $Pr_6O_{11}$ (Macklin, 99.9%) is preheated at 1000 °C, then stoichiometric amounts of $Pr_6O_{11}$ and $Ni(NO_3)_2 \cdot 6H_2O$ (Aladdin, 99.99%) were dissolved in citric acid solution and followed by the addition of citric acid (Alfa Aesar, 99.9%) as chelating agents. The solution was heated and stirred till the precursors dissolved to form a light green solution at 150 °C. After evaporating the water, the green nitrate gel was heated at 300 °C for 2 h. The gel was self-ignited to yield a fluffy black product. The product was pressed into pellets and heated at 1050 °C in flowing oxygen for 4 days with once intermediate grinding and pelletizing.

The phase purity and crystal structure of the obtained powders were studied on a powder XRD diffractometer (Empyrean) at room temperature. The high-pressure in-situ synchrotron X-ray diffraction (HPXRD) experiments on $Pr_4Ni_3O_{10-\delta}$ powder were conducted at room temperature using the BL15U1 beamline at the Shanghai Synchrotron Radiation Facility (SSRF). The incident X-ray beam was monochromatized to the wavelength of 0.6199 Å. A diamond anvil cell (DAC) with 200 μm culet anvils was utilized to achieve high pressure. Silicone oil was chosen as the pressure-transmitting medium. The Dioptas software was employed to analyze the diffraction patterns[43].

The HPXRD spectra were subsequently refined using the Fullprof program[44].

Electrical transport properties of $Pr_4Ni_3O_{10-\delta}$ powder under high pressure were carried out using a miniature DAC constructed from a Be–Cu alloy. The DAC utilized diamond anvils with culets ranging from 200 to 250 μm and an insulating gasket chamber about 60 to 80 μm in diameter, composed of a cubic boron nitride and epoxy mixture. The four-probe van der Pauw technique was applied for the electrical measurements, executed with KBr as the pressure-transmitting medium. In all high-pressure experiments, the pressure was jointly monitored through the high-frequency edge of the diamond phonon[45–47]. Magnetic and electronic transport properties were measured on a physical property measurement system (Quantum Design).

## 3 Results

As shown in Fig. 1(a), the crystal structure of $Pr_4Ni_3O_{10-\delta}$ can be described by the stacking of perovskite $(PrNiO_3)_3$ blocks and rock salt PrO layers along the $c$ axis. The crystal structures show a relatively low symmetry due to distortions of the $NiO_6$ octahedra. Figure 1(b) displays the results of Rietveld refinement for $Pr_4Ni_3O_{10}$ polycrystalline samples with the monoclinic space group $P2_1/a$ (space group No. 14, $Z = 4$). The obtained lattice parameters are $a = 5.3759(2)$ Å, $b = 5.4658(2)$ Å, $c = 27.572(1)$ Å, and $\beta = 90.333(2)°$. The lattice parameters agree with the previous reports for polycrystalline samples[42].

Figure 1(c) shows the temperature dependence of resistivity for the $Pr_4Ni_3O_{10-\delta}$ polycrystalline sample. The resistivity decreases as temperature cools down, indicating the nature of metal behavior. A significant jump in resistivity occurs at about 158 K, consistent with opening an energy gap at the Fermi level as the emergent spin and charge density waves. An upturn in resistivity can be observed below ~18 K. This feature may be associated with developing local moments of $Pr^{3+}$ ions Kondo scattering. Figure 1(d) shows the temperature dependence of susceptibility of $Pr_4Ni_3O_{10-\delta}$. There is no bifurcation between zero-field-cooling (ZFC) and field-cooling (FC) susceptibility curves down to 1.8 K. The susceptibility above 160 K follows the Curie-Weiss law: $\chi(T) = \frac{C}{T-\theta_{CW}} + \chi_0$, yielding $C = 6.176$ emu K/mol, $\theta_{CW} = -28.99$ K, and $\chi_0 = 3.03 \times 10^{-3}$ emu/mol. The experimentally determined magnetic moment of $Pr^{3+}$ ion is 3.51 $\mu_B$, close to the theoretical value of $\mu_{eff} = 3.58$ $\mu_B$ per $Pr^{3+}$ ion[42]. This suggests the negligible magnetic moment carried by Ni ions. The negative value of $\theta_{CW}$ indicates the antiferromagnetic nature of interactions between the $Pr^{3+}$ moments. In addition to the anomaly at 157.6 K, a small peak in the derivative of susceptibility can be observed at 4.3 K, which is related to the antiferromagnetic transition of $Pr^{3+}$. To further study the ground state of $Pr_4Ni_3O_{10-\delta}$, we performed isothermal magnetization measurements at different temperatures, as shown in Fig. 1(e). The magnetization increases linearly as the magnetic field increases, revealing antiferromagnetic correlations. The derivative of magnetization at 2 K shows two broad peaks at 1.1 and 5.3 T field, respectively, suggesting magnetic-field-induced magnetic transitions. The magnetic transitions are absent above 4 K, further confirming that the magnetic order develops at low temperatures.

Figure 2(a) displays the synchrotron powder XRD patterns measured on $Pr_4Ni_3O_{10-\delta}$ polycrystalline sample under 5 to 75 GPa pressures. The peaks shift under pressure, while no new reflection peaks appear. As the evolution of the (1 1 7) and (2 2 0) peak positions against pressure plotted in Fig. 2(b), a change of the slope can be observed at about 20 GPa, indicating a structural phase transition. The XRD patterns above 20 GPa in Fig. 2(c) can be well refined by the tetragonal *I4/mmm* rather than the monoclinic *P2$_1$/a*, confirming the structural transformation that resembles the structural transition in La$_4$Ni$_3$O$_{10-\delta}$[20]. Figures 2(d) and 2(f) show the evolution of lattice parameters as pressure [See supplementary table S1]. The lattice constants and cell volume show a monotonical decrease under pressure, *a* and *b* tend to be uniform near the structural phase transition. Pressure dependent evolution of the volume (*V*) follows the third-order Birch-Murnaghan equation, yielding bulk modulus $B_{0,\text{LP}}$ = 206.4 GPa and $B_{0,\text{HP}}$ =141 GPa for the low- and high-pressure phases, respectively. The different $B_0$s also evident a structural phase transition under high pressure. Figure 2(f) displays the tetragonal structure of the high-pressure phase, wherein distortions of NiO$_6$ octahedra can be negligible, and the interlayer Ni–O–Ni angle changes from 165.1(1)° to 180°.

We measured several resistances on Pr$_4$Ni$_3$O$_{10-\delta}$ polycrystalline samples under various pressures, as shown in Fig. 3 and Fig. S1. The resistance initially shows a weak insulating behavior with a rapid upturn at 175 K and 0.8 GPa in Fig. 3(a). The upturn in resistance at 175 K should be related to the intertwined density wave transition. While this transition temperature is higher than that observed from resistance at ambient pressure, $T_{\text{DW}}$[42]. The weak anomaly in resistance at ~4 K is consistent with the formation of antiferromagnetic order in the Pr sublattice (Figure S1). As pressure increases, the density waves transition temperatures are reduced, and the metallicity is enhanced. Figure 3(b) shows a drop in resistance at 23.6 GPa, which becomes more pronounced with increasing pressure. The resistance of Pr$_4$Ni$_3$O$_{10-\delta}$ under pressure is reminiscent of La$_4$Ni$_3$O$_{10-\delta}$, suggesting a superconducting transition. The $T_c$ is increased by pressure to a maximum of 30 K at 54.1 GPa. An upturn in resistance can be observed below 5 K, possibly arising from the Kondo scattering of the Pr ion. Figures 3(c) and 3(d) present the *R*(*T*) curves at 41.8 and 54.1 GPa, respectively, under various magnetic fields. The increasing magnetic fields progressively suppress the resistance anomaly, confirming that the anomaly corresponds to a superconducting transition. Figure 3(e) shows the upper critical field, $\mu_0H_{c2}$, fitting using the empirical Ginzburg-Landau formula $\mu_0H_{c2}(T) = \mu_0H_{c2}(0) (1 - t^2)/(1 + t^2)$, where $t = T/T_c$. The fitted upper critical field $\mu_0H_{c2}(0)$s are 21 and 29.8 T for 41.8 and 54.1 GPa, respectively.

A pressure-dependent density wave and superconductivity phase diagram is displayed in Fig. 3(f). In the lower pressure region, Pr$_4$Ni$_3$O$_{10-\delta}$ undergoes two transitions, forming the intertwined SDW and CDW on the Ni sublattice and the antiferromagnetic order on the Pr sublattice. The density wave transition of Ni is apparently enhanced and then progressively suppressed with increasing pressure. The structure transforms from *P2$_1$/a* to *I4/mmm* at ~20 GPa, followed by a superconducting transition. Such a coincident transition behavior suggests that the structural transition is a prerequisite to emerging superconductivity, which is analogous to the bilayer nickelate La$_3$Ni$_2$O$_7$ and trilayer nickelate La$_4$Ni$_3$O$_{10}$. Upon further increasing the pressure, a dome-shaped superconducting region with a maximum $T_c$ of 30 K at 54.1 GPa is obtained. The superconductivity persists up to 68.8 GPa that is the highest pressure measured in this measurement.

In summary, Pr$_4$Ni$_3$O$_{10-\delta}$ polycrystalline samples were synthesized using the sol-gel method. At ambient pressure, Pr$_4$Ni$_3$O$_{10-\delta}$ develops density waves on the Ni- and Pr-sublattices at about 158 and 4.3 K, respectively. Under high pressure, Pr$_4$Ni$_3$O$_{10-\delta}$ undergoes a structural transition from the monoclinic *P2$_1$/a* space group to the tetragonal *I4/mmm* space group at around 20 GPa. Accomanied by the structural transition, signatures for superconductivity are observed with a maximum $T_c$ of 30 K. The observation of superconductivity in Pr$_4$Ni$_3$O$_{10-\delta}$ suggests that high-temperature superconductivity is not limited at La-Ni-O nickelate systems; rare-earth engineering is a vital strategy to explore high-temperature superconductivities in the RP phase of nickelates.

Acknowledgments. This work was supported by the National Key Research and Development Program of China (Grants No. 2023YFA1406000, 2023YFA1406500), the National Natural Science Foundation of China (Grants No. 12425404, 12474137, and 12174454), the Guangdong Basic and Applied Basic Research Funds (Grants No. 2024B1515020040, 2024A1515030030), Shenzhen Science and Technology Program (Grant No. RCYX20231211090245050), Guangzhou Basic and Applied Basic Research Funds (Grant No. 2024A04J6417), and Guangdong Provincial Key Laboratory of Magnetoelectric Physics and Devices (Grant No. 2022B1212010008).

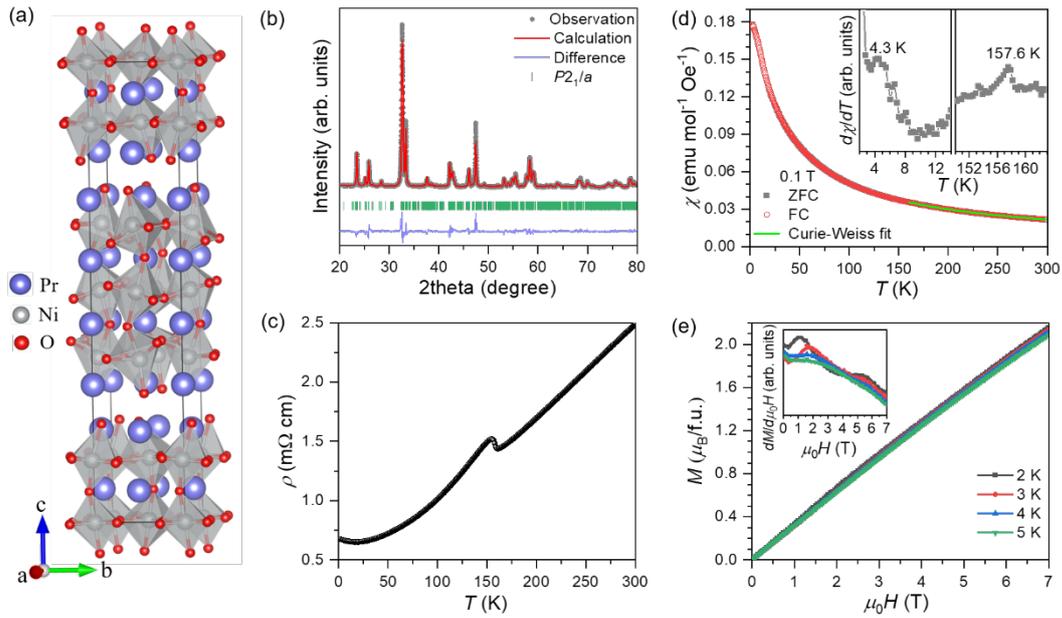

Fig. 1. (a) Crystal structure of $Pr_4Ni_3O_{10-\delta}$. (b) Powder x-ray diffraction patterns and its Rietveld fitting results of $Pr_4Ni_3O_{10-\delta}$ polycrystalline sample. (c) Temperature-dependent resistivity curve of $Pr_4Ni_3O_{10-\delta}$ from 1.8 to 300 K at ambient pressure. (d) Temperature dependence of zero-field-cooling (ZFC) and field-cooling (FC) magnetic susceptibility under a magnetic field of 0.1 T for $Pr_4Ni_3O_{10-\delta}$. The inset shows the derivative of ZFC susceptibility. (e) Magnetic field dependence of magnetization at different temperatures and the inset displays their derivative curves.

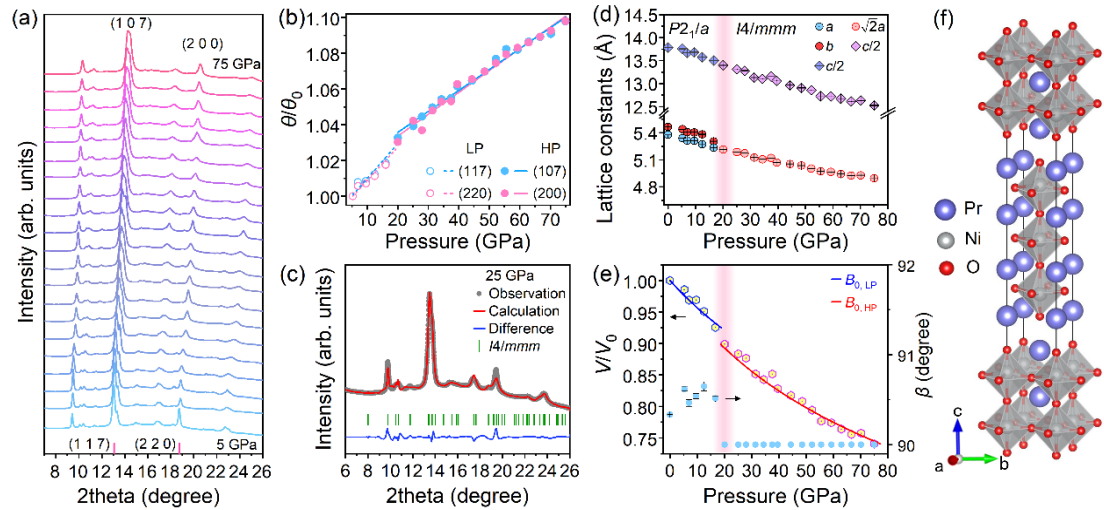

Fig. 2. (a) Room-temperature synchrotron powder XRD patterns of pressurized $Pr_4Ni_3O_{10-\delta}$ up to 75 GPa. (b) Evolution of the (117) and (220) peak positions under pressure and the fitted solid and dashed lines show the different variation trends of low- and high-pressure phases. (c) Rietveld refinement results of synchrotron powder x-ray diffraction data for $Pr_4Ni_3O_{10-\delta}$ at 25 GPa. (d) Lattice constants $a$, $b$, and $c$ extracted from the Rietveld refinement results of the synchrotron-based XRD. (e) Pressure dependence of the cell volume and the refined Ni-O-Ni angle along the c axis ($\beta$) under pressure obtained from the Rietveld refinements. The solid lines indicate the evolution of the cell volume as a function of pressure fitted by the third-order Birch-Murnaghan equation. (f) Crystal structure of $Pr_4Ni_3O_{10-\delta}$ at 25 GPa.

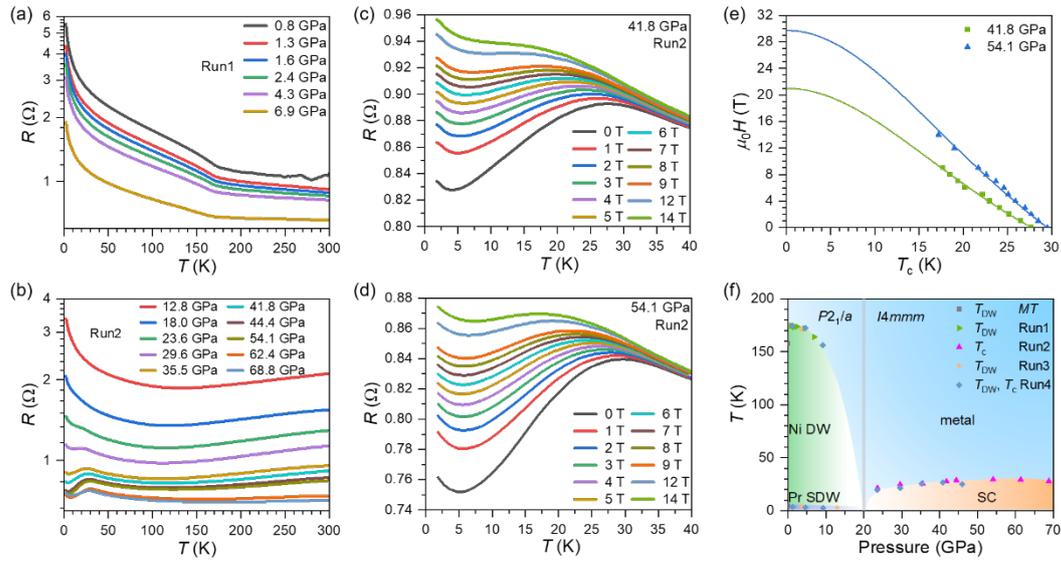

Fig. 3. (a) Resistance of $Pr_4Ni_3O_{10-\delta}$ as a function of temperature from 0.8 to 6.9 GPa in run 1. (b) Temperature-dependent resistance of $Pr_4Ni_3O_{10-\delta}$ as a function of temperature from 12.8 to 68.8 GPa in run 2. Temperature-dependent resistance measured at (c) 41.8 GPa and (d) 54.1 GPa under various magnetic fields in run 2. (e) The Ginzburg–Landau fittings of the upper critical fields at 41.8 and 54.1 GPa. (f) A temperature-pressure phase diagram of $Pr_4Ni_3O_{10-\delta}$.

Supporting Information

# Signature of Superconductivity in Pressurized Trilayer-nickelate Pr$_4$Ni$_3$O$_{10-\delta}$

Xing Huang[1#], Hengyuan Zhang[1#], Jingyuan Li[1], Mengwu Huo[1], Junfeng Chen[1], Zhengyang Qiu[1], Peiyue Ma[1], Chaoxin Huang[1], Hualei Sun[2*], Meng Wang[1*]

[1]*Center for Neutron Science and Technology, Guangdong Provincial Key Laboratory of Magnetoelectric Physics and Devices, School of Physics, Sun Yat-Sen University, Guangzhou, Guangdong 510275, China*
[2]*School of Sciences, Sun Yat-Sen University, Shenzhen, Guangdong 518107, China*
# These authors contribute equally to this work.
*sunhlei@mail.sysu.edu.cn (H. L. Sun), wangmeng5@mail.sysu.eud.cn (M. Wang)

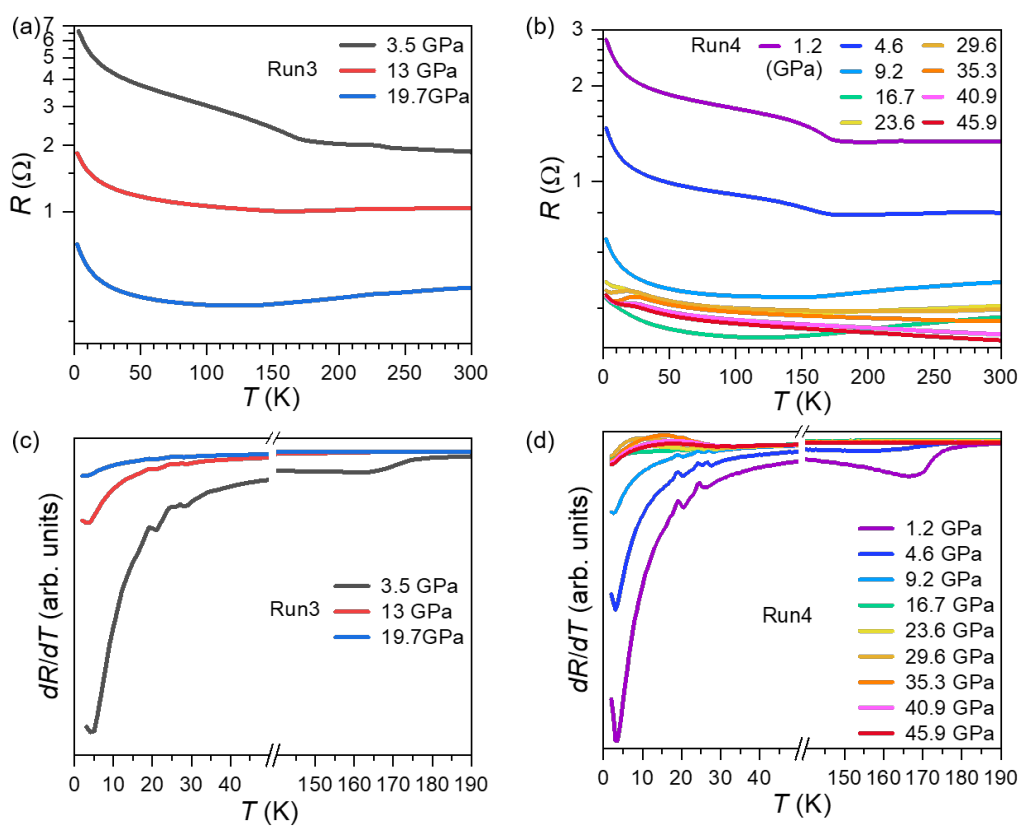

Fig. S1. (a) Resistance of Pr$_4$Ni$_3$O$_{10-\delta}$ as a function of temperature from 3.5 GPa to 19.7 GPa in run 3. (b) Electrical resistivity of Pr$_4$Ni$_3$O$_{10-\delta}$ as a function of temperature from 1.2 GPa to 45.9 GPa in run 4. The derivative curves of resistance in (c) run 3 and (f) run 4.

**Table S1** Crystallographic parameters of pressurized $Pr_4Ni_3O_{10-\delta}$ obtained from Rietveld refinements of high-pressure in-situ synchrotron X-ray diffraction (HPXRD). Silicone oil was chosen as the pressure-transmitting medium.

| Pressure (GPa) | a(Å) | b(Å) | c(Å) | β(°) | Space group |
|---|---|---|---|---|---|
| 0 | 5.3759 | 5.4658 | 27.5723 | 90.33 | $P2_1/a$ |
| 5.28 | 5.3387 | 5.4378 | 27.5100 | 90.62 | $P2_1/a$ |
| 6.99 | 5.3122 | 5.4059 | 27.3389 | 90.46 | $P2_1/a$ |
| 9.57 | 5.3104 | 5.4042 | 27.3593 | 90.54 | $P2_1/a$ |
| 12.47 | 5.2721 | 5.3824 | 27.1531 | 90.65 | $P2_1/a$ |
| 16.66 | 5.2346 | 5.3045 | 27.0046 | 90.51 | $P2_1/a$ |
| 20.06 | 3.6859 | 3.6859 | 26.8002 | 90 | $I4/mmm$ |
| 25.12 | 3.6669 | 3.6669 | 26.6201 | 90 | $I4/mmm$ |
| 27.93 | 3.6568 | 3.6568 | 26.5587 | 90 | $I4/mmm$ |
| 31.39 | 3.6239 | 3.6239 | 26.2651 | 90 | $I4/mmm$ |
| 34.37 | 3.6095 | 3.6095 | 26.2033 | 90 | $I4/mmm$ |
| 37.53 | 3.6203 | 3.6203 | 26.3139 | 90 | $I4/mmm$ |
| 39.56 | 3.5842 | 3.5842 | 26.1253 | 90 | $I4/mmm$ |
| 44.40 | 3.5736 | 3.5736 | 25.9304 | 90 | $I4/mmm$ |
| 48.53 | 3.5609 | 3.5609 | 25.8159 | 90 | $I4/mmm$ |
| 52.20 | 3.5374 | 3.5374 | 25.7117 | 90 | $I4/mmm$ |
| 55.56 | 3.5159 | 3.5159 | 25.4539 | 90 | $I4/mmm$ |
| 59.32 | 3.5101 | 3.5101 | 25.4508 | 90 | $I4/mmm$ |
| 62.92 | 3.4945 | 3.4945 | 25.3437 | 90 | $I4/mmm$ |
| 66.66 | 3.4831 | 3.4831 | 25.2323 | 90 | $I4/mmm$ |
| 70.18 | 3.4848 | 3.4848 | 25.2709 | 90 | $I4/mmm$ |
| 75.03 | 3.4605 | 3.4605 | 25.0548 | 90 | $I4/mmm$ |